# A Game-Theoretic Model of Human Driving and Application to Discretionary Lane-Changes

Jehong Yoo, *Member, IEEE* and Reza Langari, Senior *Member, IEEE*

*Abstract*— In this paper we consider the application of Stackelberg game theory to model discretionary lane-changing in lightly congested highway setting. The fundamental intent of this model, which is parameterized to capture driver disposition (aggressiveness or inattentiveness), is to help with the development of decision-making strategies for autonomous vehicles in ways that are mindful of how human drivers perform the same function on the road (on which have reported elsewhere.) This paper, however, focuses only on the *model development* and the respective *qualitative assessment*. This is accomplished in unit test simulations as well as in bulk mode (i.e. using the Monte Carlo methodology), via a limited traffic micro-simulation compared against the *NHTSA 100-Car Naturalistic Driving Safety* data. In particular, a qualitative comparison shows the relative consistency of the proposed model with human decision-making in terms of producing qualitatively similar proportions of crashes and near crashes as a function of driver inattentiveness (or aggressiveness). While this result by itself does not offer a true quantitative validation of the proposed model, it does demonstrate the utility of the proposed approach in modeling discretionary lane-changing and may therefore be of use in autonomous driving in a manner that is consistent with human decision making on the road.

## I. INTRODUCTION

Understanding human driving behavior in ordinary traffic has been a crucial topic in traffic safety and efficiency [1, 2]. To this end, researchers have developed a number of models that are meant to capture the behavior of drivers on the road [3-5]. These studies have revealed a number of factors, including driver intent and disposition (e.g. aggressiveness/inattentiveness) as well as traffic setting and roadway conditions that exert strong influences on the driver reaction and subsequently impact his/her decision-making process. In particular, drivers choose lanes, vary their speed and maintain headway based on their level of aggressiveness/inattentiveness and their perception of the surrounding vehicles' behaviors [6].

In this context, there are two main aspects to human driving: *car-following* in the longitudinal direction [7, 8] and *lane-changing* in multi-lane situations [5, 9]. To this end, researchers have provided a range of car-following models whose goal has been to produce an adequate representation of the acceleration characteristics of vehicles [3, 7, 8]. For instance, the well-known Intelligent Driver Model (IDM) produces acceleration as a continuous function of the subject vehicle's velocity, velocity difference relative to the lead vehicle, and current and desired headway [3]. Recent works have attempted to adapt these models to reflect certain human attributes, such as delayed recognition of the visual field and neuromuscular response of drivers [10, 11].

Lane-changing behavior has also been actively studied by a growing number of researchers [9, 12-14]. These works have focused on revealing the heuristics of lane-changing as the basis to create effective traffic flow modeling tools as well as to develop autonomous driving strategies. Gipps [12], for instance, has defined a set of factors that cause the driver to change lanes, including headway, speed and proximity to the desired exit ramp. Ahmed [9], among others, has utilized a gap acceptance model to assess if an adjacent gap is acceptable for lane-changing purposes. Likewise, Hidas [13] has proposed the notion of *driver courtesy*, a type of cooperation among drivers, in modeling lane-change and merging operations. Kim [14] has proposed a modified IDM to deal with conflicts among a group of vehicles when there is not a sufficient gap to merge and, in this respect, is similar to the forced merging model or Hidas's driver courtesy scheme.

In addition, Swaroop and Yoon have developed an emergency lane-change maneuver in response to the presence of obstacles within the overall framework of vehicle platooning [15] while Jula et al. have performed an analysis on the kinematics of lane-changing maneuvers and presented a minimum longitudinal spacing criterion to avoid crashes [16]. Kanaris and Ioannou have also proposed a certain *minimum safety* spacing for lane-changing and merging in automated highway systems [17].

These models all aim at a safe lane-changing approach. However, *aggressiveness* and *inattentiveness* are also important factors that affect driver behavior and subsequently impact traffic safety [18, 19]. In this context, we realize that aggressiveness and inattentiveness are not exactly the same factors. Neither do we claim that inattentiveness and the broader issue of driver distraction are the same factors. Indeed

J.-H. Yoo was with Texas A&M University, College Station, TX 77843 (email: jehong.yoo@tamu.edu).

R. Langari is with Texas A&M University, College Station, TX 77843 (email: rlangari@tamu.edu).

distraction can be considered a temporal mode of behavior that while potentially serious, once the source of distraction is removed, the driver returns to a normal state. In this paper, inattentiveness is considered a *dispositional* mode that, as with aggressiveness, is sustained over a longer period. As noted above, however, we do not claim that the two are precisely the same dispositions. We believe that in certain instances inattentiveness and aggressiveness may manifest themselves similarly, however, as when an inattentive/aggressive driver changes lanes irrespective of limited spacing in the target lane. There are also instances where these factors do not produce similar behaviors as for instance in maintaining headway. However, even in this case the behaviors may be similar in certain instances as in aggressive tailgating vs. maintaining a small headway by simply being inattentive. All said, one could argue that inattentiveness may be viewed in certain respect as a purpose-less aggressiveness (or behaviorally similar to it albeit with less intent.) In fact, while medical facts are not necessarily certain, certain studies relate inattentiveness and aggressiveness, although the relation of these studies to driver behavior modeling is not fully established [20].

Aside from aggressiveness/inattentiveness, a number of researchers have studied other factors that influence the driving behavior (such as gender and age) based on observational data [21, 22]. However, existing human driver models that are utilized in both traffic flow modeling and in autonomous driving have generally assumed *nominal* behavior as well as *reasonable* perception-action models, which may not be uniformly held across the spectrum of drivers and driving conditions [4, 23].

With this in mind, in the present work we consider a somewhat broader range of driving behaviors, particularly variations in aggressiveness/attentiveness or limitations of perception (mainly in terms of assessment of visual field), towards building a driver decision model with some utility in autonomous driving as noted in our related work [24] and further outlined in the concluding section of the paper. The *qualitative* validation that is offered in this work, however, focuses on a set of limited traffic simulations, which provide a better venue to illustrate the performance of the model and offer an intuitive assessment of it. The purpose of this assessment or validation process is not, however, to show the efficacy of the approach in exact quantification of the impact of driver aggressiveness or inattentiveness (or limitations of visual perception) on the traffic flow. Rather, this qualitative assessment or validation is intended to offer a basis for using the proposed approach in developing intuitively reasonable decision logics for autonomous driving, realizing that such logics do need to be validated rigorously in their own setting as our continued work on this topic is presently focused on.

*A. Game Theory*

Since its inception by Borel in the 1920s [25, 26] and subsequent works by Von Neumann and Morgenstern as well as by Nash [27, 28], game theory has been used as a reasonable model of decision-making in many areas of social science (particularly economics) and engineering [27-34]. In particular, Fisk [35] has pointed out that two behavioral models from game theory can be used in transportation modeling, as for instance in intercity travel and signal optimization. Optimality in this context is evaluated on the basis of payoffs resulting from the decisions by (and interactions among) the respective participants [36]. The Nash game delivers the optimal solution in non-cooperative games in general [29]. On the other hand, the Stackelberg game guarantees the best payoff for every player when there exists a hierarchical structure among players, namely when players are divided into a *leader*, who has the power to choose his/her strategy first, and a *follower*, who should choose his/her action after the leader's decision has been made [30, 31]. In this context we should point out that the Stackelberg model does not require a physical precedence order (although this may indeed be the case in driving), only that there is a play order in the sense that the lead player does have a priority in action as is the case in discretionary lane-changing in that, typically, one driver considers making a lane change in view of the current traffic state and must assess what the others would do in response and subsequently take action at which point others will slow down to allow the driver to change lanes or maintain their speeds/accelerate to prevent such a lane change. We shall discuss this issue further in the later sections of this paper.

*B. Game Theory in Transportation*

Game theory has been applied in various ways to study the effects of *policy*, *decisions*, and/or the *actions* of individual agents in transportation system. These studies can be broadly classified into two categories: *infrastructural* regulation studies (traffic control problem) and *agent-oriented* studies (vehicle placement or route decisions). In the first category, researchers have employed game theory on dynamic traffic control or assignment problems. For instance, Chen and Ben-Akiva [37] adopted a non-cooperative game model to study the interaction between a traffic regulation system and traffic flow to optimally regulate the flow on a highway or an intersection while Li and Chen [33] addressed the ramp-metering problem via Stackelberg game theory. Su et al. [38] have also used game theory to simulate the evolution of a traffic network.

In the second category vehicles are regarded as game participants and traffic rules are generally considered to be implicit in the respective decision models [32, 34, 39]. In this context Kita [40] has worked to address the merging-giveway interaction between a through car and a merging car as a two-person non-zero sum non-cooperative game. This approach can be regarded as a game theoretic interpretation of Hidas' *driver courtesy* scheme [6] from the viewpoint that the vehicles share the payoffs or heuristics of the lane-changing process. This leads to a reasonable traffic model although the approach does not address the uncertainties resulting from the actions of other drivers. Moreover, one cannot guarantee that the competing vehicle would act as determined by the game solution since that vehicle may be able to consider other factors that the subject vehicle cannot take into account.

In recent studies Talebpour et al [41, 42] have considered the notion of *incomplete information* as part of the game formulation process and have developed a model that in certain respects addresses the aforementioned concern. Likewise, Altendorf and Flemisch [43] have developed a game theoretic model that addresses the issue of *risk-taking* by drivers and its impact on the traffic flow. Their study focuses on the cognitive aspects of this decision making process and its impact on traffic safety. Likewise, Wang, et al. [44] have used a *differential game* based controller to control a given vehicle's car-following and lane-changing behavior while in [45], the authors have applied an Iterative Snow-Drift (ISD) game on *cross-a-crossing* scenario. In [46] Stackelberg game theory was used solve conflicts in shared space zones while in [47] the authors compared heavy vehicle and passenger car lane-changing maneuvers on arterial roads and freeways. Note that (Elvik [48] offers a rather complete review of related works in this area but only up to 2014 while the literature in this area continues to evolve.)

In summary, the studies listed above have worked to address the problem of contention in driving particularly as it relates to lane-changing and its impact on the traffic flow. The present work is inline with these studies but provides a rather detailed description of driver behavior while incorporating a simplified vehicle dynamic model, and attempts to show the utility of game theory in modeling the types of behaviors that impact traffic flow. More relevant to the current debate in autonomous driving, this model offers a potential approach to implementing an intuitive decision logic that is consistent with the behavioral aspects of ordinary human drivers and therefore offers a potentially sensible approach to implementing naturalistic autonomous driving strategies for use in mixed-traffic where the behavior of autonomous vehicles must remain consistent with common driving norms.

*C. Premise and Contribution of the Current Work*

In this paper, we develop an individual driver decision model based on Stackelberg game theory, which is pertinent to discretionary lane-changing. The premise here is that, the sequential structure of decision making in discretionary lane-changing is best represented by the Stackelberg game theory where the driver attempting the lane change has the *lead* role in terms of initiating the chain of decisions that may subsequently involve other vehicles (to give way, to compete and/or to execute subsequent lane changes of their own). In the present work, this theory is combined with a simplified vehicle dynamic model to develop a simulation model to evaluate the impact of driver decisions in discretionary lane-changing. (Mandatory lane-changing such as in highway merging is considered in a separate work [49].) Moreover, we pay attention to unsafe outcomes that can be caused by the driver's behavioral dispositions such as aggressiveness or inattentiveness. This is done in the definitions of the so-called *payoffs* associated with the game theoretic approach.

The model is intuitively validated in unit tests and via Monte Carlo simulations, which correlate the possibility of collision with the level of aggressiveness/inattentiveness of drivers as well as inter-vehicular distances. This validation step has been the basis for the use of this model in developing an intuitive decision logic for autonomous vehicles in mixed traffic as reported elsewhere [24], and may also support driver education campaigns and transportation policy analysis.

*D. Organization of the paper*

This paper is organized as follows: Section II describes the configurations of the problem at hand. Section III gives a formulation of traffic situation via Stackelberg game theory, discusses utility design for simulating driver behavior, and presents the solution of the game theoretic formulation. Section IV offers unit test results, which are organized as a collision possibility model. A limited traffic flow simulation is subsequently performed to qualitatively assess the effectiveness of the collision possibility model in bulk mode. In Section V a summary of our results is provided and conclusions are drawn with respect to further development of the proposed model.

## II. SYSTEM CONFIGURATION

The traffic situation considered here consists of three components: the traffic regulation system, the traffic setting, and the vehicle itself, which includes a driver. In this work, normal traffic rules such as speed limits are assumed to exist although these are not subject to game theoretic analysis. (They are implicitly considered in the modeling process and incorporated in the simulation studies.) A simple perception model captures the driving setting in terms of the positions of surrounding vehicles in view of the driver's perceptual limits (perhaps affected by age or other factors) and reflected in the decision logic and in the decision logic. The driver's decision model or logic i.e. where to go or which vehicle to follow, is based on Stackelberg game theory. This model does not generate the actual vehicle motion trajectories (be it in the same lane or during lane-changes.) Smooth trajectories are separately generated in view of physical limits of the vehicle (e.g. acceleration/deceleration rates) while steering and speed are managed via proportional plus derivative controllers. The vehicle dynamic model in turn translates these inputs into vehicle performance variables, namely velocity, position, and the yaw rate of the vehicle.

*A. Perception model*

This function classifies the given vehicle's surroundings into *leading* and *following* vehicles, according to their longitudinal positions in each lane. In reality, vicinity recognition can degenerate due to internal and/or external conditions. In this paper, we focus on the *internal* factors, although external conditions such as weather can also be incorporated in the modeling process. At present, however, these are not explicitly considered but may be to a certain degree lumped under the overall perceptual limit factor discussed shortly. To this end, we add artificial errors to approximate the uncertainty in recognizing the surroundings. This is described in more detail shortly.

*B. Vehicle Control*

The goal of this function is to precisely execute the driver's decision, be it to maintain lane position or to conduct a lane-change. Subsequent to a simple smooth trajectory generation process, two proportional plus derivative (PD) controllers are used to reproduce the driver's low-level controls of headway or speed as well as steering angle in consideration of the lane-change trajectory. Once again, both longitudinal and lateral controllers are limited by physical factors (acceleration limits, cornering stiffness values) and additional parameters that express the driver's disposition. That is to say, an aggressive (or in certain instance an inattentive) driver will may produce more drastic longitudinal and lateral accelerations than a normal or cautious driver will. Both controllers are tuned to be slightly over-damped to avoid set-point oscillations. This does not preclude the possibility of vehicle dynamic instability (say rollover) should drastic maneuvers be executed. However, even under most aggressive cases, the generated lane change trajectories do not lead to rollover, largely due to the range of parameters selected for vehicle characteristics in this study.

### C. Vehicle Dynamics

We consider a two-wheel vehicle model as depicted in Figure 1.

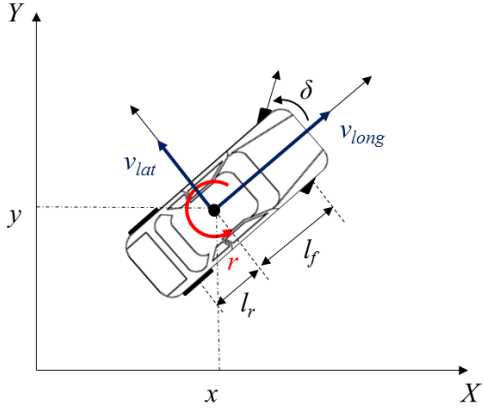

**Figure 1. Planar view of vehicle in motion.**

The vehicle model is given as

$$\begin{bmatrix} \dot{v}_{lat} \\ \dot{r} \end{bmatrix} = \begin{bmatrix} -\frac{C_f + C_r}{mv_{long}} & \frac{-l_f C_f + l_r C_r}{mv_{long}} - v_{long} \\ \frac{-l_f C_f - l_r C_r}{I_z v_{long}} & -\frac{l_f^2 C_f + l_r^2 C_r}{I_z v_{long}} \end{bmatrix} \begin{bmatrix} v_{lat} \\ r \end{bmatrix} + \begin{bmatrix} \frac{C_f}{m} \\ \frac{l_f C_f}{I_z} \end{bmatrix} \delta \quad (1.)$$

$$\begin{bmatrix} \dot{x} \\ \dot{y} \\ \dot{\theta} \end{bmatrix} = \begin{bmatrix} v \cdot \cos \\ v \cdot \sin \\ r \end{bmatrix} \quad (2.)$$

where $r$ is the yaw rate, $v_{lat}$ the lateral velocity of the vehicle, $m$ the mass of the vehicle, $I_z$ the moment of inertia, $v_{long}$ the longitudinal velocity, $\theta$ the vehicle heading, $l_f$, $l_r$ the distances between the front/rear wheel and its center of mass, and $C_{\alpha f}$, $C_{\alpha r}$ the front/rear cornering stiffness values. Other variables are evident in the figure.

### D. Driver's Manipulation of the Vehicle

To generate proper acceleration/deceleration and steering angle, we use two independent controllers for longitudinal and lateral control of the vehicle in a manner that we believe to be consistent with human driving. Both controllers (Proportional plus Derivative or PD) are used to control the relative velocity or the headway and the steering angle. The longitudinal control output, the resultant acceleration, is determined by the weighted mean of the two controls of the relative velocity and the headway. Both control outputs are limited by physical constraints of the vehicle and the driver's *disposition* (say aggressiveness as further elaborated below). In particular, an aggressive/inattentive driver may move more drastically in the longitudinal and lateral directions than a normal or cautious driver. Thus, the vehicle acceleration $a$ is given by

$$a = \min(K_{pg} \cdot e_{v,d} + K_{dg} \cdot \dot{e}_{v,d}, g_l, g_{pl}) \quad (3.)$$

where $e_{v,d}$ is the error between the reference velocity and the velocity of the vehicle or the error between the reference relative distance and the relative distance between the given vehicle and the vehicle ahead, $K_{pg}$ is the proportional gain of the longitudinal controller, $K_{dg}$ is the derivative gain of the longitudinal controller, $g_l$ is the acceleration limit that can be changed by the driver's disposition, and $g_{pl}$ is the physical limitation of acceleration or deceleration. Likewise, the steering angle is defined by

$$\delta = \min(K_{pl} \cdot e_{lat} + K_{dl} \cdot \dot{e}_{lat}, \delta_{lat}, \delta_{pl}) \quad (4.)$$

where $e_{lat}$ is the error between the reference lateral position and the lateral position of the vehicle, $K_{pl}$ is the proportional gain of the lateral controller, $K_{dl}$ is the derivative gain of the longitudinal controller, $\delta_{lat}$ is the steering angle limit that can be altered by the driver's disposition, and $\delta_{pl}$ is the physical limitation of the steering angle. Here, $\delta_{lat}$ can be obtained by using the following ratio which is known as lateral acceleration gain [50]:

$$\frac{a_{yl}}{\delta_{lat}} = \frac{v^2}{57.3Lg + K_{us}v^2} \quad (5.)$$

where $a_{yl}$ denotes the lateral acceleration limit that can be changed by the driver's *disposition* (again, say aggressiveness as further discussed in the sequel), $L$ is the wheelbase, $K_{us}$ is the understeer gradient of the vehicle, and $g$ is the acceleration of gravity.

### E. Collision detection

Vehicles are assumed as rectangles that have certain widths and lengths. Thus, in the simulations, collision between any two vehicles can be detected in terms of the overlapping area between the projections of these rectangles using the Separating Axis Theorem [51] as depicted in Figure 2.

To formulate an index $I_{col} \in [0,1]$ to represent the collision possibility, we use the gaps between two rectangles along the separating axis.

$$D_{proj(v,i)} = \begin{cases} \min\left(\left|\frac{v_i \cdot v}{v_i}\right|\right) & if \quad \forall \frac{v_i \cdot v}{|v_i|} < 0 \\ \min\left(\left|\frac{v_i \cdot v}{v_i}\right| - |v_i|\right) & if \quad \forall \frac{v_i \cdot v}{|v_i|} > |v_i| \\ 0 & otherwise \end{cases} \quad (6.)$$

where $D_{proj}$ denotes the gaps along the separating axis, $v_i$ a vector defining the rectangle, and $v$ the vector to the opposite corner. We the define the collision possibility index, $I_{col} \in [0,1]$, as the exponential inverse of the gap.

$$I_{col} = e^{-\sqrt{\frac{D_{col,v}^2 + D_{col,u}^2}{2}}} \quad (7.)$$

where

$$D_{col,v} = \sqrt{D_{proj(v,1)}^2 + D_{proj(v,2)}^2}. \quad (8.)$$

$D_{col,u}$ is defined in a similar manner. Note that the collision possibility index is 1 when two rectangles overlap and 0 when the gap between them approaches infinity.

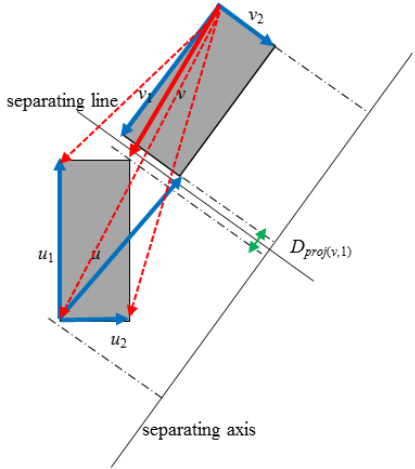

**Figure 2. Schematic view of Separating Axis Theorem.**

### III. GAME THEORETIC FORMULATION

*A. Game Definition*

We configure a straight road with three lanes as the smallest meaningful traffic setting for this purpose. This setting offers three basic choices: changing lane to the left, going straight, and changing lane to the right. No lane preference is assumed although this can be incorporated in the model. We assume the road to be occupied by two kinds of vehicles: vehicles that incorporate decision makers or have *intentions* and vehicles that follow given set paths at set speeds. The latter vehicle act as props and construct the boundary of the simulation.

*B. Game Formulation*

In the present study we formulate a game with three players as depicted in Figure 3: the vehicle itself, serving as the *lead* vehicle and the two *follower* vehicles in the two adjacent lanes. This implies that the subject vehicle does not attempt to influence the vehicles ahead but does survey the presence of these vehicles for decision-making purposes. The Stackelberg game is therefore defined as a three-person finite game with three levels of hierarchy:

*Players* : $P1, P2, P3$

*Strategy space* : $S_{st} = L_1 \times \Gamma_1 \times \Gamma_2 \times \Gamma_3$

$L_1 = \{1,2,3\}, \Gamma_{1,2,3} = \{L, S, R\}$

where $P1$ designates the leader in the specific game (with its boundaries defined in Figure 3 instance), $P2$ the nearest follower (which may also act as a leader in a subordinate game involving $P2$ and $P3$) and $P3$ as the second follower to $P1$ (or as the direct follower to $P2$, should $P2$ consider a lane change to the right as part of its subordinate game with $P1$; $L_1$ the lane number of the leader, and $\Gamma_{1,2,3}$ the actions: i.e. going left, $L$, going straight, $S$, and going right, $R$.

Moreover, the game is considered to be *dynamic*, in the sense that the payoffs corresponding to the respective strategies change as a function of the driving situation. In other words, and as we shall see shortly, the payoffs or utility functions are functions of the inter-vehicular distances and velocities and as such dynamically change. However, the decision to initiate a lane-change depends on these payoffs, and in turn reflects the traffic state (including the respective velocities) at the time the lane-change decision is made.

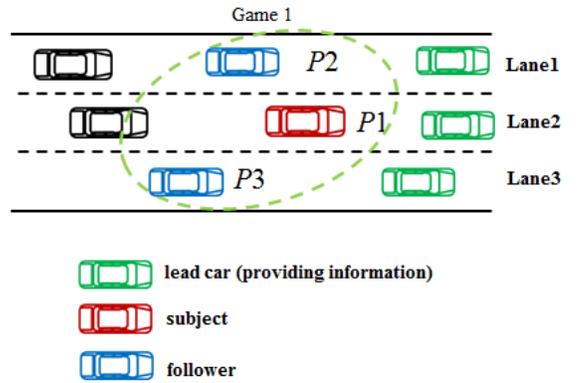

**Figure 3. Schematic of 3-player Stackelberg game.**

We do assume that vehicular velocities do not radically and abruptly change over the short temporal horizon in which the respective lane-change occurs. This presents a certain limitation of our approach, which ignores intentional or unintentional abrupt changes in the behaviors of the drivers (beyond what is expected based on their presumed level of aggressiveness/inattentiveness), and further reflects the fact that our attribution of aggressiveness/inattentiveness is dispositional and not temporal. In other words, temporally short behavioral changes are not considered in this work; neither are true distracted driving behaviors that may occur over a short time-frame.

We should also point out that in our work (and to our knowledge different from most, if not all related studies) players do not share their payoff matrices. This implies that each player may have a different perspective on the game in which s/he is involved, and each player may perceive its optimal strategy in a way that may or may not be exactly the same as the way in which other players view theirs. This also means that any other vehicles may also consider the subject vehicle as a player in a different game and respond to the subject vehicle's actions in that context. While this adds certain level of realism to the proposed approach, it also implies that attributed best decision by one player may indeed be different from that which is perceived (for the same player) by others. This fact itself implies a more subtle theoretical nuance vis-à-vis the notion of game equilibrium as it is traditionally attributed to the Stackelberg game model, and which requires further investigation. However, as we have stated at the outset, our study's emphasis has been on developing an intuitively realistic model as we work to explore the theoretical basis for variations from non-standard formulation of the Stackelberg game theory.

*C. Utility design*

We define two complementary utility functions related to two of the factors that Gipps [12] has considered; *speed advantage* and *unacceptable collision risk*. These utility functions incorporate the driver aggressiveness, via an index, $q$, as an important element that affects driving behavior and traffic safety [22]. If the driver is completely aggressive or inattentive, the index $q$ is 1 (or 100%.) If the driver is normal, $q$ is 0.5 (or 50%), and if the driver is completely cautious, or attentive, $q$ is 0 (0%.) As noted here, the use of an aggressiveness index is also intended to reflect the human driver's inattentiveness as well as willfully aggressive driving [52-54].

*1. Utility associated with headway*

Since the focus of this work is on discretionary lane-changing behavior, we assume that drivers wish to maintain an appropriate *headway* while moving at their desired *speed*. Thus we define $U_h$ as:

$$U_h = \min(d_r, \alpha(q) \cdot d_v) \quad (9.)$$

where $d_r$ is the headway distance (relative distance between the subject vehicle and the vehicle ahead), $d_v$ denotes the visibility distance for a normal driver; $\alpha(q)$ modifies $d_v$ as a function of driver aggressiveness/inattentiveness. There is no unique choice in this case as attribution of aggressiveness or inattentiveness may lead one to use a decreasing or increasing function $\alpha(q)$ based on context. Aggressive or inattentive driver may tailgate persistently requiring a decreasing $\alpha(q)$ or, as least in the case of some aggressive drivers, may demand a larger headway, leading to an increasing $\alpha(q)$. Aside from this point, this utility measure can be quantified relative to the current lane or relative to any target lane to the left or right side of the current lane as further discussed below.

*2. Utility reflecting lane-change*

We include a utility function that reflects the feasibility of a lane-change in view of possible collision with a competing vehicle in the target lane, $U_l$ is defined as

$$U_l = d_r - v_r \cdot T(q) - D_{suf} \quad (10.)$$

where $d_r$ denotes the relative distance between the subject and competing vehicles, $v_r$ the respective relative velocity, $T$ the prediction time as a monotonically decreasing function of $q$ and $D_{suf}$ the distance required to conduct the lane-change: i.e. a multiple of the diagonal length of the subject vehicle.

*3. Total utility*

For each player we define the total utility as:

$$U = U_h + U_l \quad (11.)$$

The utility of staying in a given lane is clearly only due to headway in that lane since $U_l$ would be zero for this case. In considering a lane-change, however, $U_l$ would generally be nonzero per the definition of this term in the previous section and possibly negative if the driver is cautious with an appropriately large prediction time, $T$. This could in principle prevent a lane change in the case of cautious drivers as it is later seen in the results section of this paper. On the other hand, for an aggressive driver, the impact of this term could be minimally negative or perhaps even positive (although in practice that is rare unless the competing vehicle in the target lane is sufficiently behind.) This would lead to a higher possibility of a lane change in this case as noted in Section IV.

*D. Driver Reaction and Limitation of Perception*

Driver reaction time and poor prediction of other vehicles' actions are crucial factors in transportation [55-58]. To consider these issues, we consider a degradation factor due to delayed recognition of the presence of other vehicles and late reaction to these. This is implemented by changing the decision maker's recognition point that is used to assess lane intrusion as depicted in Figure 4.

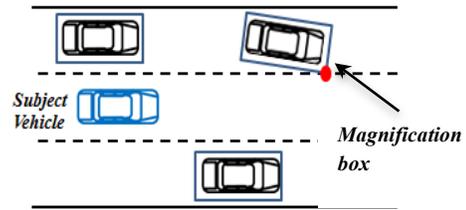

**Figure 4. Boundary recognition response.**

It is assumed that the magnification ratio of the rectangle around the protruding vehicle would vary with respect to aggressiveness/inattentiveness. In addition, if the driver is attentive, the lane transition is identified when the closest front corner of the vehicle meets the set criterion. If the driver is inattentive/aggressive, it is assumed that the recognition point is near the planar center of the protruding vehicle. In other words, a cautious/attentive driver is mindful of lane intrusions while an aggressive or inattentive driver "barrels" through, as it were, less concerned about the potential risk of collision.

*E. Solution of the Stackelberg Game*

Clearly, appropriate action should be chosen among the final utility pairs in order to simulate the driver's behavior. Since every vehicle follows the Stackelberg game, the solution ($\gamma^{1*}$, $\gamma^{2*}$, $\gamma^{3*}$) of the game is obtained by the following 3-person Stackelberg equilibrium equations [59] with the designed utility values,

$$\min_{\gamma^2 \in S^2(\gamma^{1*})} \min_{\gamma^3 \in S^3(\gamma^{1*};\gamma^2)} U^1(\gamma^{1*},\gamma^2,\gamma^3) = \max_{\gamma^1 \in \Gamma^1} \min_{\gamma^2 \in S^2(\gamma^1)} \min_{\gamma^3 \in S^3(\gamma^{1*};\gamma^2)} U^1(\gamma^1,\gamma^2) \stackrel{def}{=} U^{1*}$$

$$S^2(\gamma^1) \stackrel{def}{=} \{\xi \in \Gamma^2 : \min_{\gamma^3 \in S^3(\gamma^1;\xi)} U^2(\gamma^1,\xi,\gamma^3) \geq \min_{\gamma^3 \in S^3(\gamma^1;\gamma^2)} U^2(\gamma^1,\gamma^2,\gamma^3), \forall \gamma^2 \in \Gamma^2\}$$

$$S^3(\gamma^1;\gamma^2) \stackrel{def}{=} \{\xi \in \Gamma^3 : U^3(\gamma^1,\gamma^2,\xi) \geq U^3(\gamma^1,\gamma^2,\gamma^3), \forall \gamma^3 \in \Gamma^3\}$$

where $U^1$, $U^2$, and $U^3$ denote the respective utilities of the leader and the follower vehicles, keeping in mind that there is a priority order among the followers as discussed in Section III.*B*; $\gamma^1$ is the possible action of the leader; $\gamma^2$ and $\gamma^3$ are respective reactions of the followers. Every two-person finite Stackelberg game admits a strategy for the leader [59], which can be extended to the three-person finite Stackelberg game with three levels of hierarchy because it can be understood as two successive two-person games given that the interaction among the leader and its first follower can place the latter in the position of being the leader in a game with the second follower.

The set of $\gamma^3$, $S^3$, is first obtained as the argument of the maximum payoff for the follower with respect to $\gamma^1$ and $\gamma^2$. In turn, $\gamma^2$ is obtained in the same manner for a given $\gamma^1$ with the consideration of the follower's reaction. Since the follower's (the second player against the first leader and the follower against the first and second leaders) payoff is *unique* but the strategy is not necessarily unique [59], we establish an order among the strategies that have the same payoffs, such that the driver chooses "going straight" if the payoffs of "going straight" and "changing lanes" are the same, and the driver in the second lane chooses the first lane when the first and third lanes offer the driver the same payoffs.

Thus, every possible $\gamma^1$ has the accompanying unique reactions of the second leader and the follower. Accordingly, $\gamma^{2*} \in S^2(\gamma^{1*})$ and $\gamma^{3*} \in S^3(\gamma^{1*}; \gamma^{2*})$ are determined as the optimal strategies of the second leader and the follower corresponding to $\gamma^{1*}$ and the pair ($\gamma^{1*}$, $\gamma^{2*}$), respectively. In simple terms, the driver in our proposed model predicts the two followers' responses and chooses the best strategy based on that prediction. This is essential to the notion of Stackelberg game theory and from our perspective applicable to discretionary lane changing. In other words, the driver considering the lane change serves as the lead decision maker or lead player while considering the actions of the competing vehicles in the adjacent lanes in selecting his/her optimal decision. Note that since every driver has his/her own Stackelberg game-based decision model and does not share the utilities, the best responses for the followers are not guaranteed. This is in important point, and as we have stated earlier, may present certain theoretical challenges, but adds realism to the modeling process.

### F. Schematic Presentation of the Game

Since the game has a strategy space that is difficult to visualize, it is necessary to design a matrix-like formulation that is composed of lanes and players; i.e. the vehicles. In Figure 5, the rows designate the hierarchy among the players and the columns show the probable lane selections. The figure depicts the case when *P*1, *P*2 and *P*3 choose *R*, *R*, and *S* respectively as their strategies and the utilities are determined by their physical longitudinal positions.

Every strategy combination is marked on the matrix by laterally changing the players' lanes. Since we defined a game that has three highway lanes, the left side, $C_L$, of the leftmost lane and the right side, $C_R$, of the rightmost lane are assumed to be areas where changing lanes is impossible, such as the centerline and the road shoulder. Therefore, vehicles have the least payoffs for the strategies that make them enter these areas. In addition, since the human driver's visibility is bounded by physical limitation, the vehicles beyond the visibility distance are not considered. Figure 5 represents one possible strategy pair in the strategy space. As we can see on the left side of the figure, the strategy pair is not likely to lead to better payoffs for *P*1, rather than the payoffs of other strategy pairs. In other words, what is sktehced here is for illustration purposes and does not necessarily reflect the final (optimal) decision of the subject vehicle.

## I. SIMULATIONS

### A. Unit Test Scenarios

We tested a specific scenario consisting of two vehicles in addition to the three front dummy vehicles that form the boundary of the simulation area as depicted in Figure 6. As stated earlier, every two-person finite Stackelberg game admits a strategy for the leader [59], which can be extended to the three-person finite Stackelberg game with three levels of hierarchy because it can be understood as two successive two-person games. As a result, and to focus on the task at hand, which is qualitative assessment or validation of the overall algorithm, we consider the simpler case shown in the figure.

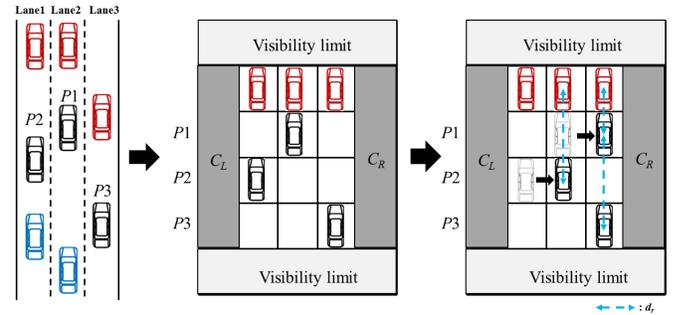

**Figure 5. Matrix-space formulation of the driving game.**

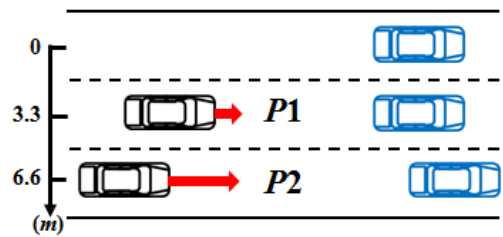

**Figure 6. Unit test scenario.**

The purpose of the scenario is to focus on the interaction between two vehicles when they use the Stackelberg game-theoretic decision model. Vehicles are assumed to travel in the direction of the *Y*-axis, and the lanes are set in direction of the *X*-axis of the coordinate frame depicted in the figure; Lane 1 (*x*=0 *m*), Lane 2 (*x*=3.3*m*), and Lane 3 (*x*=6.6*m*). There are two initial settings for the vehicles. We set the initial positions, velocities, and the drivers' dispositions for every vehicle to determine the impact of the driver's disposition in conjunction with the design of the aforementioned utility functions. One vehicle in the third lane, among the front prop vehicles, is located slightly ahead in order to bring about a lane-change for the subject vehicle in each case. Vehicles 1 and 2 are initially in the second and third lanes, respectively. In order to test the lane-change situation when there is a fast approaching vehicle, Vehicle 2 is located 50 *m* behind Vehicle 1, but with a higher velocity. This is stated in Table 1.

Four aggressiveness combinations are tested for Vehicles 1 and 2. These represent the interactions of two normal drivers, an aggressive driver and a cautious driver, two aggressive drivers, and two cautious drivers, in that order.

**Table 1. Initial positions of test scenarios**
**(x0: initial lateral, y0: initial longitudinal.)**

|  | Initial Conditions | | |
| --- | --- | --- | --- |
|  | x0 (m) | y0 (m) | v0 (m/s) |
| Vehicle1 | 3.3 | 0 | 100 |
| Vehicle2 | 6.6 | -50 | 130 |

*B. Unit test results*

The unit test results are shown in Figure 7 through Figure 10. The *two-normal-drivers* test case, depicted in Figure 7 and the *two-aggressive-drivers* test case, depicted in Figure 9 show that the driver of Vehicle 1 in the second lane changes its lane to the third lane, which initially has a larger space in front, and then Vehicle 2 changes its lane to the second lane because its free space is now restricted by Vehicle 1. While the pattern of these transitions is similar, Figure 9 shows that the aggressive drivers' lane-changes happen sooner than the normal drivers' as depicted in Figure 7. This leads to a distinct rise in the collision possibility index (shown in solid blue) relative to the threshold (shown in a dashed red line.)

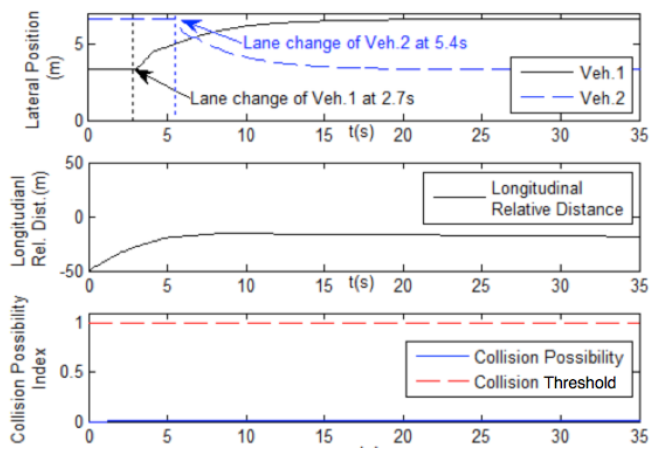

**Figure 7. Unit test for two normal drivers**

In Figure 8 the cautious driver of Vehicle 2 does not try to overtake the aggressive driver and maintains a safer relative distance. In Figure 10, with the combination of two cautious drivers, no driver changes lanes.

In all, the most dangerous instant appears in Figure 9. The reason for this result is that the more aggressive/inattentive the drivers are, the less mindful they are of the presence of vehicles in the target lane. Conversely, if one driver is cautious, even though the other is aggressive, as in Figure 8, their collision possibility remains relatively low.

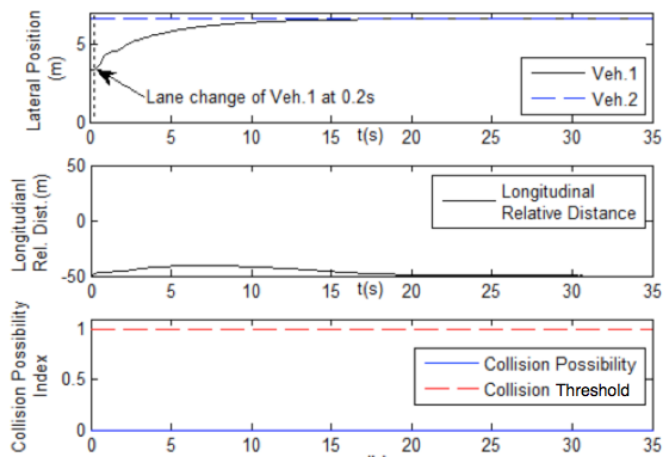

**Figure 8. Unit test for aggressive and cautious drivers**

While these appear to be intuitively valid, they do not tell the whole story as it were since in more congested setting the interactions can be correspondingly more complex as discussed below.

*C. Monte Carlo Simulation*

To determine the general effects of aggressiveness/inattentive driving, we performed a Monte Carlo simulation involving randomized longitudinal positions and constructed a model to estimates the collision possibility as a function of longitudinal positions and aggressiveness/inattentiveness of drivers. We tested 100 cases with random longitudinal inter-vehicular distances. The longitudinal positions of Vehicles 1 and 2 are uniformly distributed in the range of 0 to 50 *m* and 0 to -50 *m*, respectively.

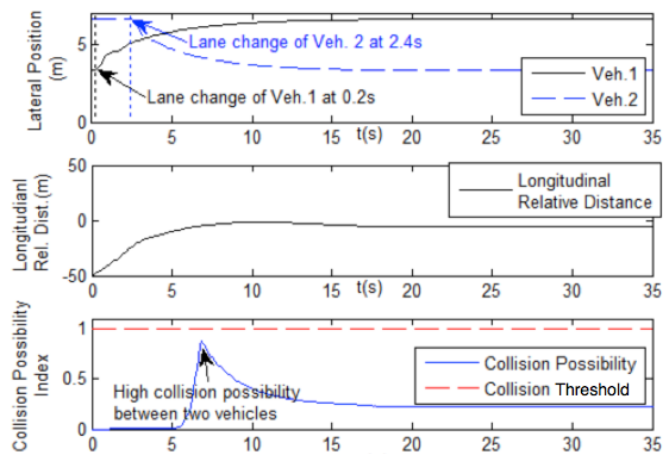

**Figure 9. Unit test for two aggressive drivers**

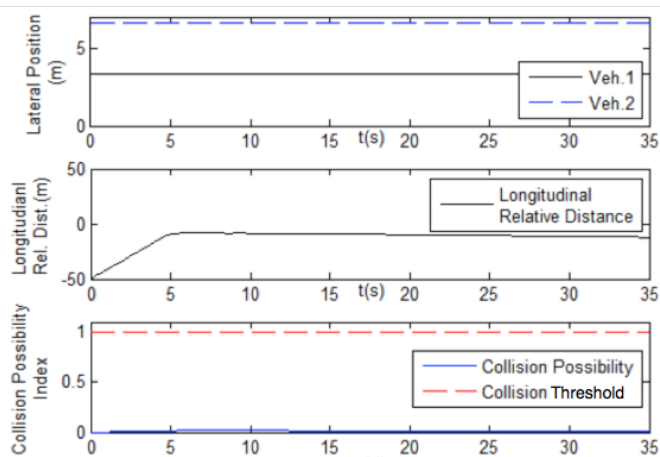

**Figure 10. Unit test for cautious drivers**

Three aggressiveness/inattentiveness combinations (Normal/Normal, Aggressive/Cautious, and Aggressive/Aggressive) were considered; the case that both drivers are cautious (or timid as denoted in the figure) led to no meaningful results. These results are summarized in Figure 11, which shows the collision possibility as a function of driver aggressiveness and inter-vehicular distance. Although the highest level of potential collision is different for every test case, the trend in collision possibility appears similar to the unit test results. This is particularly evident by looking at the right side of Figure 11 (inter-vehicular distances larger than 50m), where for normal/normal and aggressive/cautious (timid) cases, the collision possibility remains low and increases when the combinations of drivers are mutually aggressive as it was in the unit test studies.

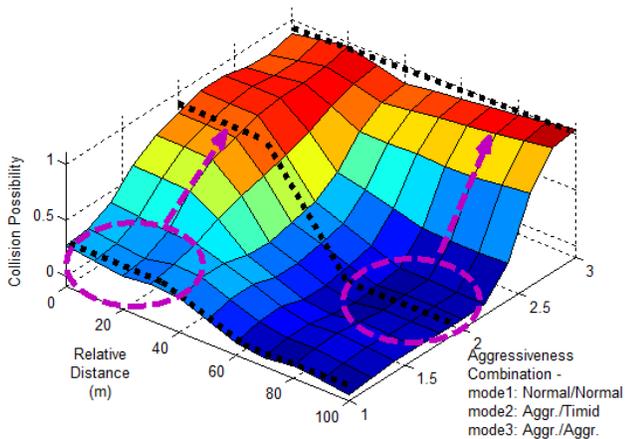

**Figure 11. Collision possibility index as a function of driver aggressiveness and longitudinal separation.**

As the inter-vehicular distances decrease, there is a general trend towards higher collision possibility even when drivers are not particularly aggressive (the left side of the graph in Figure 11.) This appears intuitively reasonable as well. It is also worth noting that even when drivers are normal, as the inter-vehicular distances decreases, there is a general increase in collision possibility as it is evident in the dotted line crossing the span of the figure. This trend is even more pronounced in the case of aggressive vs. cautious drivers in ways that was not evident in the unit-test studies shown earlier. The explanation here is that as the inter-vehicular distances decrease, an aggressive driver may continue to engage in a rapid lane change but the preceding cautious (timid) driver may not react, leading to limited spacing between the vehicles in the same lane, and producing a high assessment of the collision possibility index. In all, the results seem to reflect the qualitative reasonableness of the model.

*D. Microscopic Traffic Simulation*

In this section, the proposed driver decision model is used as the basis for a microscopic traffic simulation. To accomplish this, we conducted a series of multiple vehicle simulations for the evaluation of crash occurrence and cumulative collision possibility according to drivers' aggressiveness/inattentiveness combinations. A 200 m section of a three-lane highway was simulated as shown in Figure 12. Vehicles enter and leave the section on a consistent basis. Thus, the density of traffic in the section is maintained. For example, if 10 groups of vehicles pass through the section over 1 minute when the density of the vehicles is 10 veh/section, the flow rate is 100 veh/min.

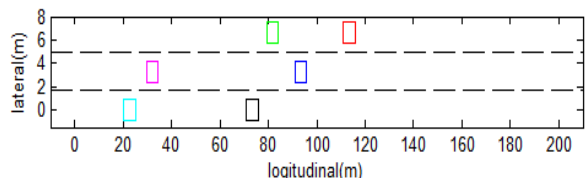

**Figure 12. Configuration of the traffic flow simulation**

As a semi-quantitative validation measure we compare the results of this microscopic simulation with the crash data from the so-called NHTSA *100-Car Naturalistic Driving Safety* dataset [2]. The NHTSA study tracked the behavior of the drivers of 100 vehicles equipped with video and other sensors for over a year and 2 million miles, producing 42,300 hours of data. The drivers of the vehicles were involved in 82 crashes, 761 near crashes, and 8,295 critical incidents.

In order to compare our results with the NHTSA study, a crash is assumed to occur in our study when the collision possibility index reaches 1. In addition, when the collision possibility index exceeds 0.5, a near crash is assumed to occur. The density of traffic is 6 veh/section for consistency with the NHTSA dataset. In order to perform a comparison we assume the inattentive case from the NHTSA study to correspond to our model's aggressiveness level of 75%. The results are listed in Table 2.

**TABLE 2. OCCURRENCE OF CRASHES IN MODEL**

| Crash type | Number of crashes |
|---|---|
| Crash (Attentive) | 1 |
| Near Crash (Attentive) | 12 |
| Crash (inattentive) | 2 |
| Near Crash (inattentive) | 26 |

In order to make sense of these and compare with the NHTSA study results, the occurrence of crashes are converted to crash rate per MVMT (Million Vehicle Miles Traveled). Nevertheless, the results of our model cannot be directly compared with the NHTSA (field) data because translation of the levels of aggressiveness and near crash possibility threshold are evidently based on our selective parametrization. However, a qualitative comparison shows a certain level of effectiveness of the model in assessing the impact of aggressiveness/inattentiveness on traffic safety as shown in Figure 13. Note in particular that the results of the model have a similar level of overapproximation to the field data across the various cases considered.

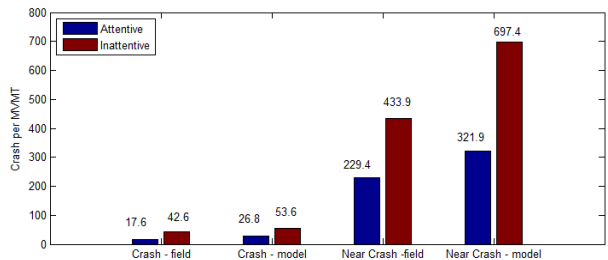

**Figure 13. Comparison of crash rate**

The cumulative collision possibility results from our microscopic simulation are presented in Figure 14. To evaluate the effects of aggressive/inattentive drivers, the ratio of aggressive drivers is set to 50% and 100% in the test cases of Aggr./Timid and Aggr./Aggr. in Figure 14. Note that 5, 50, 500 runs of the simulation with the density of 6 veh./section yield 30, 300, and 3000 vehicle simulations respectively.

Likewise, 40, 400 and 4000 vehicle simulations result from a density of 8 veh/section. Bars in each figure show the difference caused by different drivers' aggressiveness combinations. It can be seen that aggressiveness combinations influence the accumulated collision possibility regardless of the number of vehicles. Similar to the previous result, the collision possibility shows a growing tendency as the ratio of aggressive drivers increases.

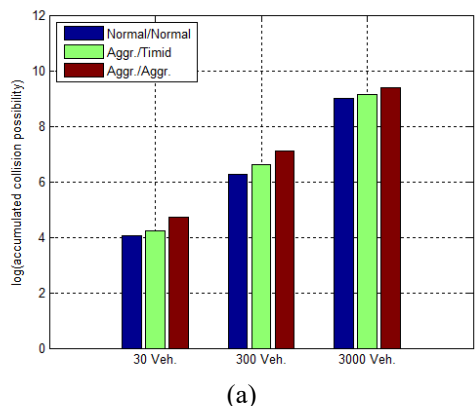

(a)

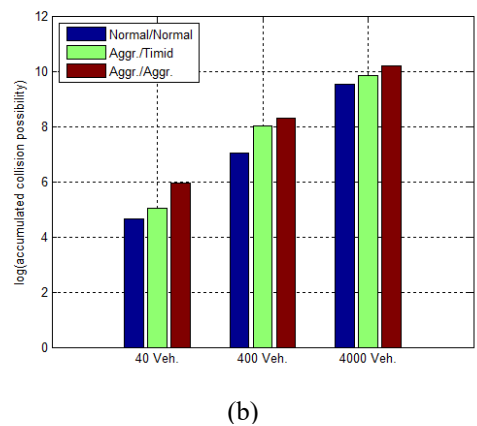

(b)

**Figure 14. Distribution of cumulative collision possibilities**

In addition, when the density of vehicles is higher in Figure 14 (b) compared with Figure 14 (a), the collision possibility increases under every aggressiveness combination, which means shorter relative distances escalate the collision possibility as also described in the previous subsection. These results in general agree with other works that have studied how driver aggressiveness and traffic density affects traffic safety [18, 22].

## II. DISCUSSION AND CONCLUSIONS

### A. Discussion

In this paper, we developed a Stackelberg game based driver reasoning/decision model for discretionary lane changing in highway driving with the consideration of driver disposition (aggressiveness/inattentiveness) and certain limitations of perception. Since Stackelberg game theory can be used to demonstrate the human decision making process when the event has the structure of a hierarchy, a game theoretic traffic simulation based on 3-person Stackelberg game theory and simplified vehicle dynamics have been presented in order to simulate the human behavior in certain driving situations (essentially slightly congested freeway setting). The game theoretic model was intended to address the typical decision-making process on the road where drivers can at best only *predict* other drivers' behaviors and *cannot control* them.

We assumed that every vehicle in the simulation was involved in a 3-person game (2-person game when appropriate),

where the given vehicle is the first leader of the game. This is a limitation of the model as it stands in that we do not directly consider the effect of multiple preceding vehicles in the competing lanes although there is a chain of games leading back from the subject vehicle so the *indirect* effect is included. This assumption appears reasonable in a light to moderately congested traffic settings but will need to be revised in more congested situations.

In order to capture the characteristics of a variety of drivers, we presented (the game-theoretic) utility (payoff) designs that reflect the drivers' intentions and are influenced by their dispositions (aggressiveness/inattentiveness). From the simulation results, we showed that Stackelberg game theory played a reasonably effective role in choosing a certain action among the possible action sets, much like human reasoning.

Second, we presented a collision possibility model in terms of the drivers' aggressiveness levels. We further showed that the above game theoretic approach to the traffic simulation could provide sufficiently explainable demonstrations. The unit tests showed plausible results about the interactions of vehicles reflecting the impact of driver aggressiveness/inattentiveness on the possibility of collision during lane-changing maneuvers.

Subsequently a compact collision possibility model was developed via a limited Monte Carlo simulation, reflecting the general result that mutually aggressive combinations can be a critical reason for high collision possibilities and smaller inter-vehicular distances magnifies the collision possibility with almost every combination of dispositions.

A limited traffic micro-simulation further showed the bulk effect of a similar trend over a 200m section of a three-lane highway with collision possibility showing an upwards trend with an increase in traffic density (as well as an increase the proportion of aggressive/inattentive drivers.) Likewise, comparison of our model with the NHTSA *100-Car Naturalistic Driving Safety* data set [2] has shown that our Stackelberg game based driver decision model can mimic (if not perfectly replicate) observation-based results from the aforementioned study that traffic safety is influenced by the driver's aggressiveness/inattentiveness.

*B. Conclusions*

As stated at the outset, our intent has been to lay the foundation for developing a reasonably intuitive model for lane-changing decisions that can be applicable to autonomous driving (or advanced driver assist systems capable of executing an autonomous lane-changing maneuver.) From the above results it does appear that, a Stackelberg game theoretic decision model is potentially of value in this context. Clearly, such a decision-making model or logic will need to be extensively validated in more elaborate simulator and test-track studies. However, our preliminary results (as reported elsewhere [24]) indicate that this is in principle possible, provided the aforementioned decision logic is augmented with an effective trajectory optimization scheme (for instance using a robust model predictive control strategy) to address the impact of uncertainty in the behavior of competing vehicles (as well as the physical limitations of the subject vehicle and roadway conditions.) This point also raises a potential concern over computational complexity of the approach vis-à-vis implementation in real-time. In this respect, Stackelberg game theory actually offers a potential benefit in that unlike the Nash game theory produces a *deterministic* solution, and computationally more tenable in real-time. Combination of game theoretic analysis with trajectory optimization does present computational challenges, however, but no more than model predictive control (a commonly proposed approach in this realm) does. All said, however, one needs to be mindful of the computational aspects of any decision logic from a practical standpoint.

There are additional limitations in the present work that need to be addressed. Among these, addressing the impact of uncertainty in perception (beyond what is included in this work) is critically important. Drivers as well as autonomous vehicles are hindered by uncertainty in assessing the true traffic state due to weather, light saturation, color effects, and the like. Including these factors will enable the simulation environment and the related decision models to be more complex but also more realistic. Likewise, assessing the true level of aggressiveness of other vehicles is critical in executing appropriate decisions. Human drivers assess other drivers' dispositions (not always accurately) through a combination of visual observations and internal models of human behavior. In an ongoing work, we have developed a learning scheme that captures aspects of this but further work is needed in this area.

Clearly, there are other extensions of the current work that can be considered, such as including lane preference (left vs. right) in lane-changing and considering the impact of vehicle platoons (vehicles in competing lanes acting in unison over certain time-windows). We are currently engaged in this analysis and will report on the work in the near future but the subject domain is open for others to consider.

Finally, this model needs to be better validated. This is important in many respects but also costly. We are presently porting the model into a dSpace simulator for more active human integration and do intend to perform test-track studies in the future although no detailed plans have been developed. Such a validation will enable fine tuning of the model and adding realism in view of actual data, and would support not only development of autonomous vehicle decision models but may also be used in developing driving simulation games that can be used in driver education campaigns or incorporated in bona fide traffic micro-simulation environments to support policy analysis in view of traffic safety.

III. REFERENCES

[1] J. A. Michon, *A critical view of driver behavior models: what do we know, what should we do?* 1985, p. 40 p.
[2] (2006). *HS-810 593, The 100-Car Naturalistic Driving Study, Phase II - Results of the 100-Car Field Experiment*.
[3] M. Treiber, A. Hennecke, and D. Helbing, "Congested traffic states in empirical observations

[4] C. C. MacAdam, "Understanding and modeling the human driver," (in English), *Vehicle System Dynamics,* vol. 40, no. 1-3, pp. 101-134, Sep 2003.

[5] A. Kesting, M. Treiber, and D. Helbing, "Enhanced intelligent driver model to access the impact of driving strategies on traffic capacity," (in English), *Philosophical Transactions of the Royal Society a-Mathematical Physical and Engineering Sciences,* vol. 368, no. 1928, pp. 4585-4605, Oct 13 2010.

[6] P. Hidas, "Modelling lane changing and merging in microscopic traffic simulation," *Transportation Research Part C,* vol. 10, pp. 351-371, 2002.

[7] P. G. Gipps, "A Behavioral Car-Following Model for Computer-Simulation," (in English), *Transportation Research Part B-Methodological,* vol. 15, no. 2, pp. 105-111, 1981.

[8] M. Treiber, A. Hennecke, and D. Helbing, "Microscopic simulation of congested traffic," (in English), *Traffic and Granular Flow'99,* pp. 365-376, 2000.

[9] K. I. Ahmed, M. E. Ben-Akiva, H. N. Koutsopoulos, and R. G. Mishalani, "Models of freeway lane changing and gap acceptance behavior," *Proceedings of the 13th international symposium on the theory of traffic flow and transportation,* pp. 501-515, 1996.

[10] R. Deborne, R. Gilles, and A. Kemeny, "Modeling Driver Adaptation Capabilities in Critical Driving Situations," in *SAE*, 2012: SAE International.

[11] T. Lee, J. Kang, K. Yi, K. Noh, and K. Lee, "Integration of Longitudinal and Lateral Human Driver Models for Evaluation of the Vehicle Active Safety Systems," in *SAE*, 2010: SAE International.

[12] P. G. Gipps, "A Model for the Structure of Lane-Changing Decisions," (in English), *Transportation Research Part B-Methodological,* vol. 20, no. 5, pp. 403-414, Oct 1986.

[13] P. Hidas, "Modelling lane changing and merging in microscopic traffic simulation," *Transportation Research Part C: Emerging Technologies,* vol. 10, no. 5, pp. 351-371, 2002.

[14] C. Kim and R. Langari, "Game theory based autonomous vehicles operation," *International Journal of Vehicle Design,* vol. 65, no. 4, pp. 360-383, 2014.

[15] D. Swaroop and S. M. Yoon, "Integrated lateral and longitudinal vehicle control for an emergency lane change manoeuvre design," (in English), *International Journal of Vehicle Design,* vol. 21, no. 2-3, pp. 161-174, 1999.

[16] H. Jula, E. B. Kosmatopoulos, and P. A. Ioannou, "Collision avoidance analysis for lane changing and merging," (in English), *IEEE Transactions on Vehicular Technology,* vol. 49, no. 6, pp. 2295-2308, Nov 2000.

[17] A. Kanaris, E. B. Kosmatopoulos, and P. A. Ioannou, "Strategies and spacing requirements for lane changing and merging in automated highway systems," (in English), *IEEE Transactions on Vehicular Technology,* vol. 50, no. 6, pp. 1568-1581, Nov 2001.

[18] F. A. Whitlock, *Death on the road: a study in social violence* (Studies in social ecology and pathology,, no. 2). London,: Tavistock Publications, 1971, pp. xii, 212 p.

[19] L. Mizell, *Aggressive driving*. 1997, p. 12 p.

[20] A. L. Thompson, B. S. Molina, W. Pelham, and E. M. Gnagy, "Risky driving in adolescents and young adults with childhood ADHD," *Journal of pediatric psychology,* vol. 32, no. 7, pp. 745-759, 2007.

[21] B. Krahé and I. Fenske, "Predicting aggressive driving behavior: the role of macho personality, age, and power of car," *Aggressive Behavior,* vol. 28, no. 1, p. 9 p., 2002.

[22] D. Shinar and R. Compton, "Aggressive driving: an observational study of driver, vehicle, and situational variables," (in English), *Accident Analysis and Prevention,* vol. 36, no. 3, pp. 429-437, May 2004.

[23] D. D. Salvucci, "Modeling driver behavior in a cognitive architecture," (in English), *Human Factors,* vol. 48, no. 2, pp. 362-380, Sum 2006.

[24] J.-H. Yoo and R. Langari, "Development of a Predictive Collision Risk Estimation Scheme for Mixed Traffic," in *ASME 2014 Dynamic Systems and Control Conference*, 2014, pp. V001T10A005-V001T10A005: American Society of Mechanical Engineers.

[25] E. Borel and J. Ville, *Applications aux jeux de hasard* (Traite du Calcul des Probabilites et de SES Applications, no. 4/2). Paris: Gauthier-Villars, 1938.

[26] E. Borel, "La théorie du jeu et les équations intégrales à noyau symétrique," *Comptes Rendus de l'Academie des Sciences,* vol. 173, pp. 1304-1308, 1921.

[27] J. Von Neumann and O. Morgenstern, *Theory of games and economic behavior*, 60th anniversary ed. (Princeton classic editions). Princeton, N.J. ; Woodstock: Princeton University Press, 2007, pp. xxxii, 739 p.

[28] J. F. Nash, "Equilibrium Points in N-Person Games," (in English), *Proceedings of the National Academy of Sciences of the United States of America,* vol. 36, no. 1, pp. 48-49, 1950.

[29] A. Kelly, *Decision making using game theory*. Cambridge (UK): Cambridge University Press, 2003, pp. X-204 p.

[30] W. F. Bialas, "Cooperative N-Person Stackelberg Games," (in English), *Proceedings of the 28th IEEE Conference on Decision and Control, Vols 1-3,* pp. 2439-2444, 1989.

[31] A. M. Colman and J. A. Stirk, "Stackelberg reasoning in mixed-motive games: An experimental investigation," (in English), *Journal of Economic Psychology,* vol. 19, no. 2, pp. 279-293, Apr 1998.

[32] S. Kalam, M. Gani, and L. Seneviratne, "Fully non-cooperative optimal placement of mobile vehicles," (in English), *2008 IEEE International Conference on Control Applications, Vols 1 and 2,* pp. 685-690, 2008.

[33] Z.-l. Li and D.-W. Chen, "A Stackelberg game approach to ramp metering and variable speed control," *The Proceedings of the 2003 IEEE International Confuence on Intelligent Transportation Systems,* vol. 2, pp. 1061-1063, 2003.

[34] E. Semsar and K. Khorasani, "Optimal control and game theoretic approaches to cooperative control of a team of multi-vehicle unmanned systems," (in English), *2007 IEEE International Conference on Networking, Sensing, and Control, Vols 1 and 2,* pp. 628-633, 2007.

[35] C. S. Fisk, "Game-Theory and Transportation Systems Modeling," (in English), *Transportation Research Part B-Methodological,* vol. 18, no. 4-5, pp. 301-313, 1984.

[36] T. Basar and G. J. Olsder, *Dynamic noncooperative game theory*. Society for Industrial Mathematics, 1999.

[37] O. J. Chen and M. E. Ben-Akiva, "Game-theoretic formulations of interaction between dynamic traffic control and dynamic traffic assignment," *Transportation Research Record,* vol. 1617, pp. p. 179-188, 1998.

[38] B. B. Su, H. Chang, Y. Z. Chen, and D. R. He, "A game theory model of urban public traffic networks," (in English), *Physica a-Statistical Mechanics and Its Applications,* vol. 379, no. 1, pp. 291-297, Jun 1 2007.

[39] P. Yan, M. Y. Ding, and C. P. Zhou, "Game-theoretic route planning for team of UAVs," (in English), *Proceedings of the 2004 International Conference on Machine Learning and Cybernetics, Vols 1-7,* pp. 723-728, 2004.

[40] H. Kita, K. Tanimoto, and K. Fukuyama, "A game theoretic analysis of merging-giveway interaction: A joint estimation model," (in English), *Transportation and Traffic Theory in the 21st Century,* pp. 503-518, 2002.

[41] A. Talebpour, H. Mahmassani, and S. Hamdar, "Multiregime Sequential Risk-Taking Model of Car-Following Behavior: Specification, Calibration, and Sensitivity Analysis," *Transportation Research Record: Journal of the Transportation Research Board,* no. 2260, pp. 60-66, 2011.

[42] A. Talebpour, H. S. Mahmassani, and S. H. Hamdar, "Modeling lane-changing behavior in a connected environment: A game theory approach," *Transportation Research Part C: Emerging Technologies,* vol. 59, pp. 216-232, 2015.

[43] E. Altendorf and F. Flemisch, "Prediction of driving behavior in cooperative guidance and control: a first game-theoretic approach," *CogSys, Madgeburg,* 2014.

[44] M. Wang, S. P. Hoogendoorn, W. Daamen, B. van Arem, and R. Happee, "Game theoretic approach for predictive lane-changing and car-following control," *Transportation Research Part C: Emerging Technologies,* vol. 58, pp. 73-92, 2015.

[45] H. Qi, S. Ma, N. Jia, and G. Wang, "Experiments on individual strategy updating in iterated snowdrift game under random rematching," *Journal of theoretical biology,* vol. 368, pp. 1-12, 2015.

[46] R. Schönauer, M. Stubenschrott, W. Huang, C. Rudloff, and M. Fellendorf, "Modeling Concepts for Mixed Traffic: Steps Toward a Microscopic Simulation Tool for Shared Space Zones," *Transportation Research Record: Journal of the Transportation Research Board,* no. 2316, pp. 114-121, 2012.

[47] K. Aghabayk, S. Moridpour, W. Young, M. Sarvi, and Y.-B. Wang, "Comparing heavy vehicle and passenger car lane-changing maneuvers on arterial roads and freeways," *Transportation Research Record: Journal of the Transportation Research Board,* no. 2260, pp. 94-101, 2011.

[48] R. Elvik, "A review of game-theoretic models of road user behaviour," *Accident Analysis & Prevention,* vol. 62, pp. 388-396, 2014.

[49] J.-H. Yoo and R. Langari, "A Stackelberg Game Theoretic Driver Model for Merging," in *ASME 2013 Dynamic Systems and Control Conference*, 2013, pp. V002T30A003-V002T30A003: American Society of Mechanical Engineers.

[50] T. D. Gillespie, *Fundamentals of vehicle dynamics*. Warrendale, PA: Society of Automotive Engineers, 1992, pp. xxii, 495 p.

[51] G. Meyer and J. Valldorf, *Advanced microsystems for automotive applications 2010 : smart systems for green cars and safe mobility*. Berlin: Springer, 2010.

[52] B. Lin and C. X. Wu, "Mathematical Modeling of the Human Cognitive System in Two Serial Processing Stages With Its Applications in Adaptive Workload-Management Systems," (in English), *IEEE Transactions on Intelligent Transportation Systems,* vol. 12, no. 1, pp. 221-231, Mar 2011.

[53] L. Malta, C. Miyajima, N. Kitaoka, and K. Takeda, "Analysis of Real-World Driver's Frustration," (in English), *IEEE Transactions on Intelligent Transportation Systems,* vol. 12, no. 1, pp. 109-118, Mar 2011.

[54] D. Sandberg, T. Akerstedt, A. Anund, G. Kecklund, and M. Wahde, "Detecting Driver Sleepiness Using Optimized Nonlinear Combinations of Sleepiness Indicators," (in English), *IEEE Transactions on Intelligent Transportation Systems,* vol. 12, no. 1, pp. 97-108, Mar 2011.

[55] T. J. Ayres, L. Li, D. Schleuning, and D. Young, "Preferred time-headway of highway drivers," (in English), *2001 IEEE Intelligent Transportation Systems - Proceedings,* pp. 826-829, 2001.

[56] M. V. d. Van der Hulst, T. Rothengatter, and T. Meijman, "Strategic adaptations to lack of preview in driving," *Transportation Research Part F,* vol. 1, pp. 59-75, 1998.

[57] W. Van Winsum, "The human element in car following models," *Transportation Research Part F: Traffic Psychology and Behaviour,* vol. 2, no. 4, pp. 207-211, 1999.

[58] W. Van Winsum and A. Heino, "Choice of time-headway in car-following and the role of time-to-

collision information in braking," (in eng), *Ergonomics,* vol. 39, no. 4, pp. 579-92, Apr 1996.

[59] T. Başar and G. J. Olsder, *Dynamic noncooperative game theory*, 2nd ed. (Classics in Applied Mathematics, no. 23). Philadelphia: Society for Industrial and Applied Mathematics, 1999, pp. XV, 519 s.

**Jehong Yoo** completed his B.Sc. and M.S. degrees in Korea and received his PhD in Mechanical Engineering from Texas A&M University in August 2014.

**Reza Langari** received the B.Sc., M.Sc., and Ph.D. degrees in mechanical engineering from the University of California, Berkeley, in 1981, 1983, and 1991, respectively. Currently, he is a professor with the Department of Mechanical Engineering, Texas A&M University, College Station.